# TOWARDS PRECISE DETECTION OF PERSONAL INFORMATION LEAKS IN MOBILE HEALTH APPS

Alireza Ardalani, Joseph Antonucci, Iulian Neamtiu
*New Jersey Institute of Technology*

**ABSTRACT**

Mobile apps are used in a variety of health settings, from apps that help providers, to apps designed for patients, to health and fitness apps designed for the general public. These apps ask the user for, and then collect and "leak" a wealth of Personal Information (PI). We analyze the PI that apps collect via their user interface, whether the app or third-party code is processing this information, and finally where the data is sent or stored. Prior work on leak detection in Android has focused on detecting leaks of (hardware) device-identifying information, or policy violations; however, no work has looked at processing and leaking of PI in the context of health apps. The first challenge we tackle is extracting the semantic information contained in app UIs to discern the extent, and nature, of personal information. The second challenge we tackle is disambiguating between first-party, legitimate leaks (e.g., the app storing data in its database) and third-party, problematic leaks, e.g., processing this information by, or sending it to, advertisers and analytics. We conducted a study on 1,243 Android apps: 623 medical apps and 621 Health&Fitness apps. We categorize PI into 16 types, grouped in 3 main categories: identity, medical, anthropometric. We found that the typical app has one first-party leak and five third-party leaks, though 221 apps had 20 or more leaks. Next, we show that third-party leaks (e.g., advertisers, analytics) are 5x more frequent than first-party leaks. Then, we show that 71% of leaks are to local storage (i.e., the phone, where data could be accessed by unauthorized apps) whereas 29% of leaks are to the network (e.g., Cloud). Finally, medical apps have 20% more PI leaks than Health&Fitness apps, due to collecting additional medical PI.

## 1. INTRODUCTION

Mobile apps collect billions of users' data each day: Android alone has in excess of 3 billion monthly active users (Samat et al., 2022). Apps designed for medical, health, or fitness purposes, are particularly important, because the data they collect contains personally identifiable information, and medical/health data. However, one major downside of apps collecting (and users providing) this data lies in the possibility of its misuse. Data is routinely sent to advertisers; stored in the Cloud where it can be accessed by malicious actors/ PI leaks are consequential: leaked PI can be sold for profit, (mis)used to identify individuals, etc. (Van Alstin et al., 2024).

Little effort has been put into understanding and exposing leaks of data that users supply via the app's UI, especially in the context of health, fitness, or medical apps. Prior work on PI leaks have used dynamic analysis (McClurg et al., 2013) but only analyzed up to 100 apps. Some efforts used differential cryptanalysis (Continella et al., 2017) or network traffic analysis (Jia et al., 2019) (Ren et al., 2016) but only examined data that left the device, without tracking sensitive data that could be exposed in device storage. One line of prior work aimed to find leaks of hardware device identifiers, such as MAC addresses, serial numbers, device location, etc. (Arzt et al., 2014). However, we aim to find leaks of personal information, or personal identifiers such as name, weight, height, date of birth, or postal (Zip) code. Another line of work on detecting sensitive input in UI (Huang et al., 2015) has included identity and weight/height information as we do, but did not look for medical information, and did not track where the information is flowing, or who is responsible for the leak (first-party vs third-party). Work that has distinguished between first-party and third-party leaks (Rahaman et al., 2021) has only focused on hardware identifiers, and not analyzed the destination of leaks, e.g., network, logs, files, local DB storage, as we do.

## 2. APPROACH

Our approach consists of two analyses. First, we run a GUI analysis that extracts and categorizes the PI collected. Second, we run a taint (information flow) analysis to determine where the PI flows (leaks).

### 2.1 GUI Analysis

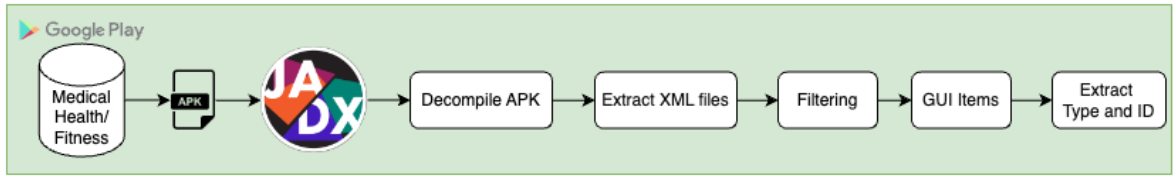

Figure 1. Overview of the GUI analysis process

**Overview.** Figure 1 shows our analysis process. Android apps are distributed as ".apk" files, so we name them APKs for short. An APK bundles app code with app resources (e.g., strings, icons, images). The dataset for this study consists of 1,243 APK files, downloaded from the Google Play Store's Medical, as well as Health&Fitness, categories. The APK files are then decompiled using JADX (JADX, 2024), yielding the source code and associated resources, including XML files.

**Extracting GUI elements.** Next, the extracted XML files, which contain the GUI layout definitions and other critical information about the application's GUI components, are processed. In Android, all GUI elements descend from class *View*. We filter out non-*View* elements, ensuring that only relevant GUI components, such as *EditText*, *Spinner*, *CheckBox* and *RadioButton*, are retained.

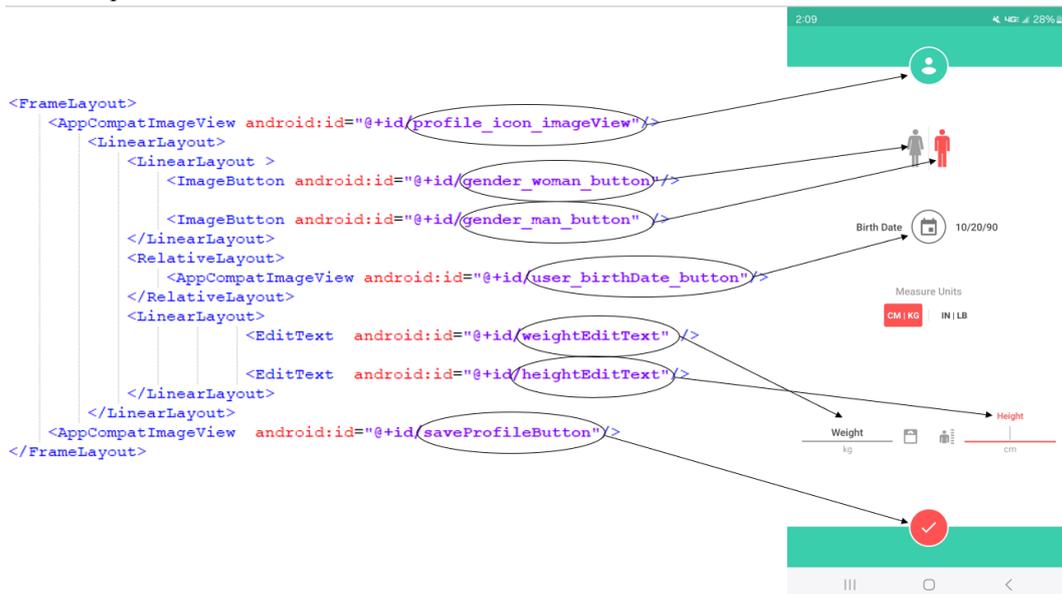

Figure 2. UI-XML mapping for app *Weight Loss Tracker*

In Figure 2 we show the mapping between XML code and UI for app *Weight Loss Tracker*. This app contains several *Views*: *EditText*, *Button*, etc. Note how in this case the *android:id* associated with UI elements facilitates analysis, because the *id* name, e.g., *id/weightEditText* or *id/user_birthdate_button*, directly encodes the PI semantics. However, there were two main additional challenges associated with UI extraction. First, there are UI elements where the *android:id* is non-suggestive, which required us to analyze other XML attributes, e.g., text, or hints associated with that *View*. Second, especially for medical information, we had to look for adjacent terms, as explained under "PI-based categorization" below.

**Personal Information (PI) definition.** Regulatory frameworks such as HIPAA in the US define "protected health information" to include health conditions, care provided, and information that can be used to identify the individual (e.g., name, phone number, social security number, birthdate, etc. (HIPAA, 2024)). The focus of this paper is on information that is collected from the user, via the GUI. Specifically, based on initial analysis of the most frequent information present in GUIs, we focus on 16 PI grouped into:

- *Identity*: email, name (first, last), address, zip code, credit card number, social security number.
- *Anthropometric/biopsychosocial*: age (birthdate), height, weight, gender.
- *Medical*: medical history, medication, blood-related, mental health, smoking or alcohol use.

**PI-based categorization**. Mapping textual information attached to UI elements onto a PI semantics was nontrivial. We used techniques from Information Retrieval along with an iterative refinement process to increase precision (gradually reduce False Positives and False Negatives). Whereas Identity PI are relatively straightforward to find, other PIs require searching for adjacent terms; in Table 1 we show some examples.

Table 1. Information retrieval: adjacent terms for PI extraction

| PI | Adjacent terms |
|---|---|
| *Medical history* | Surgery, Allergy, … |
| *Medication* | Prescription, Dosage, Dose, Drug, … |
| *Blood* | Glucose, Cholesterol, Oxygen, Pressure, … |
| *Mental health* | Stress, Panic, Anxiety, Depress, … |

## 2.2 Leak Analysis

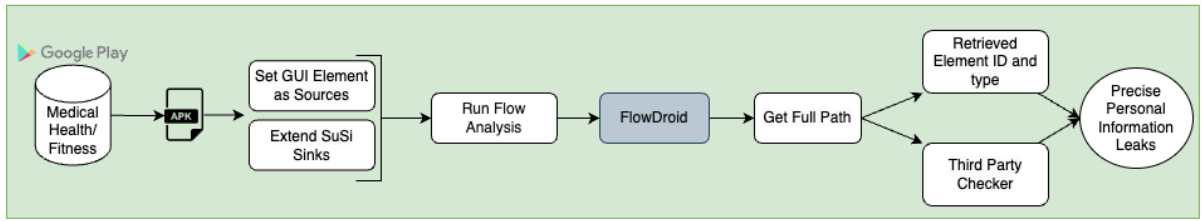

Figure 3. Overview of the leak (information flow) tracking process

**Flow analysis.** Figure 3 shows the pipeline we use, leveraging flow analysis, a standard program analysis that tracks flow of data from a *source* (origin) to a *sink* (destination). We define a *leak* as a path from a source to a sink; when multiple paths exist from one source to one sink, we only count the shortest path. Prior work on security has typically set trusted device identifiers as sources, and untrusted destinations (network, SMS) as sinks. However, in our approach, we set UI components as *sources*, while as *sinks*, we set any method that can potentially leak user information, e.g., methods that send data to the network, or write to local storage. To create the list of sink methods, we started with the SuSi list (widely-used in security research (SuSi, 2024)) as our baseline and expanded it by manually adding more method signatures, tripling the count of the baseline. As flow analyzer, we used FlowDroid (Arzt et al., 2014), which requires specifying Java methods as sinks and sources. However, in Android, UI elements are specified as XML objects, not as Java methods. Therefore, setting UI elements as sources required addressing this challenge, as follows: we used the JADX tool to map object identifiers in XML to their Java creation code in the Android-specific *R.java* file, and from there we set the *findViewById* method as a source (*findViewById* connects a UI element to its corresponding code section).

**First-party vs. third-party flows.** Another crucial aspect of precise leak attribution is distinguishing between first-party and third-party code. First-party code is identified by the package name. For example, in app *com.gotokeep.yoga.intl.apk*, code whose package is *com.gotokeep.\** or *com.gotokeep.yoga.\** is considered first-party, whereas code in packages *io.branch.\** or *com.facebook.\** is considered third-party. Of course, code in package *com.facebook.\** would be considered first-party in the Facebook app itself.

## 3. RESULTS

**App dataset**. We ran our approach on 1,243 Android apps: 623 from Google Play's Medical category and 621 from Google Play's Health&Fitness category. We used two criteria for including apps in our analysis: (1) a minimum popularity threshold ($\geqq$ 500 installs) and (2) the apps had to be in English.

### 3.1 What is the prevalence of PI collection?

Figure 4 shows the prevalence of PI collection. We found that *Email* is by far the most collected information, with 44% of Health&Fitness apps ('Fitness' for short), and 39% of Medical apps collecting it, respectively. Next, we found that *Age, Name (first, last), Phone number, Address, Gender, Height*, are collected by about 15%–20% of the apps. As expected, the collection of specialized medical information (e.g., *current medications, medical history, use of smoking/alcohol*) is more prevalent in medical apps. However, we were surprised to see that fitness apps have a comparatively higher prevalence of collecting identity information such as *email, weight, age, gender, height*, compared to medical apps.

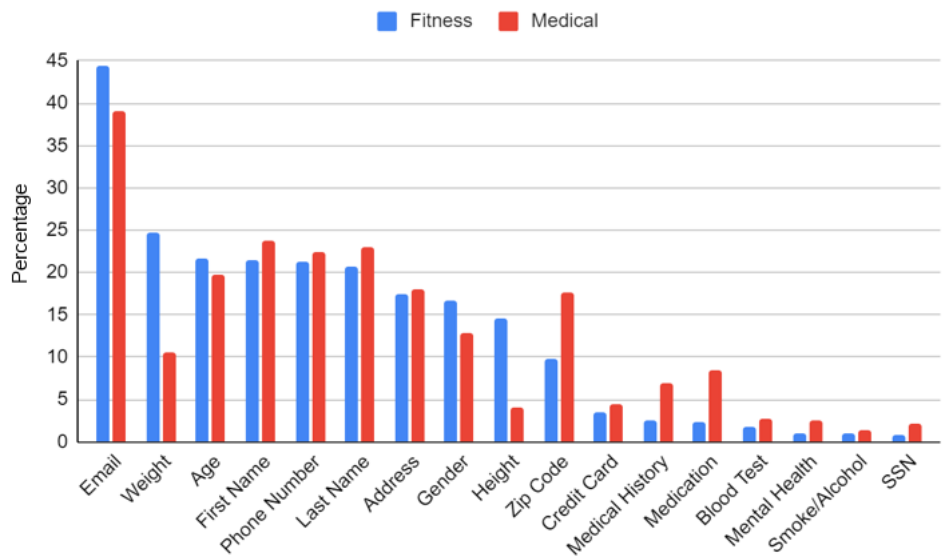

Figure 4. Prevalence of PI collected

### 3.2 First-party vs. Third-party leaks

We begin with two motivating examples that illustrate the difference between first- and third-party leaks.

    **Motivating example: first-party leak.** The mental health app *Panic Shield* (*com.panic.shield*) is designed to protect users from panic attacks. The list of user phobias, labeled fear8name, is among the PI it collects; the app collects this information via an *EditText* and stores this information in the local *SharedPreferences* database. The result of flow tracking, that starts at the source and ends at the sink, indicates the methods and VM registers involved in the flow:

```
                                                                          SOURCE
$r1=virtualinvoker0.<com.p...CreateH'chy:android.view.View findViewById(int)> ⇒
$r23 = r0.<com.panic.shield.exposure.CreateHierarchy: java.lang.String l> ⇒
Interfaceinvoke$r25.<android.cont.SharedPs$Editor:android.cont.SharedP$Editor
putString(java.lang.String,java.lang.String)>("fear8name", $r23)            SINK
```

We consider this a *first-party leak* since all the code involved belongs to the app (or Android itself).

**Motivating example: third-party leak.** The *Keep Yoga* fitness app (*com.gotokeep.yoga.intl*) collects the user's gender (encoded as a boolean) and uses the *io.branch* third-party library, which will store the gender in the local *SharedPreferences* database. The result of flow tracking is:

```
                                                                                    SOURCE
r11=virtualinvk$r6.<android.app.Activity:android.v.View findViewById(int)>($i0) ⇒
$r1 = <io.branch.ref.PrefHelper: io.branch.ref.PrefHelper prefHelper_> ⇒
putBoolean(java.lang.String,boolean)>($r1, $z0)                                     SINK
```

Note how in the flow from collecting the user's gender (source) to storing it in the database (sink), there is a method from the *io.branch* third-party library. This library is primarily used for deep linking and user engagement (Table 3). The presence of third-party methods, such as those from *io.branch*, makes this *third-party flow*. Note how third-party flows increase the risk of data leakage: the data, once handled by third-party code, can be stored locally but also potentially shared with third-parties (Reardon et al., 2019). Therefore, this second leak scenario, involving the sharing of data with third-parties, is potentially more dangerous due to the additional risk of exposing personal information to external entities.

Table 2. Leak statistics

|         | First-party | Third-party |
|---------|-------------|-------------|
| Median  | 1           | 5           |
| Average | 13.46       | 39.42       |
| Max     | 320         | 2,856       |

Table 2 shows statistics: the typical app has 1 first-party leak and 5 third-party leaks (the average values are affected by apps that have a large number of leaks, up to 320 first-party, and 2,856 third-party, as can be seen in the last row). We believe that these results are concerning, especially from the standpoint of third-party leaks. *When the ratio of leaks induced by third-party libraries to an app's own code is 5:1, essentially the developer has long lost control over who collects user data, and how this data is processed or sent.*

**Overall leaks.** Figure 5 shows histograms for the overall number of leaks, as well as first-party vs third-party leaks. The histogram on the left shows that 266 apps have less than 20 leaks. However, please note that, to be included in the figure, an app had to have at least one leak. The histogram shows that dozens upon dozens of apps have a substantial number of leaks: 62 apps have 20–40 leaks, 63 apps have 40–60 leaks, and *52 apps have more than 100 leaks*. These numbers indicate that a high number of leaks is not an isolated incident; this provides an impetus to find and analyze apps with a high number of leaks.

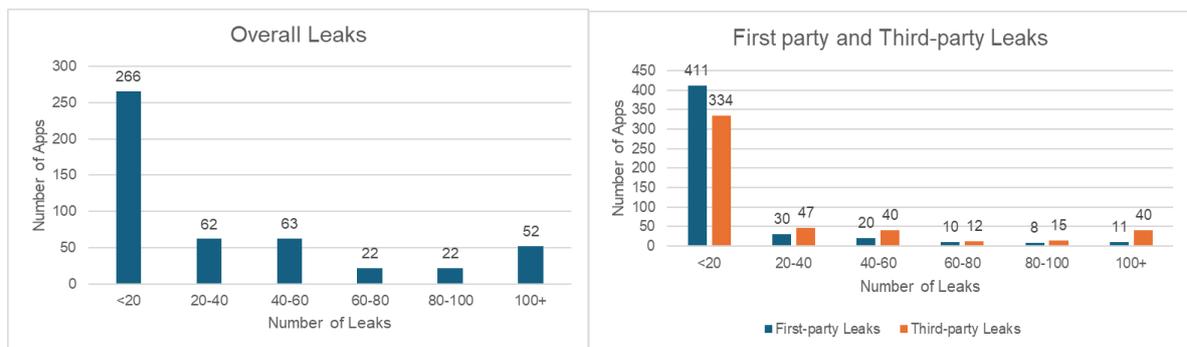

Figure 5. Leak distributions: overall (left) and first- vs -third party leaks (right)

**Most prevalent third-party libraries.** In Table 3 we show the nature and prevalence of top third-party libraries. For example *com.facebook* (which implements analytics and social media integration services) appears in 87 apps; *io.fabric*, which provides analytics, crash reporting, and user engagement services, appears in 38 apps. Note that an app may contain more than one of these third-party libraries, and there is a lot of fragmentation. While, for space reasons, we only present the top-9 libraries, there are dozens of other third-party libraries that we found in apps.

Table 3. Most frequent third-party libraries and their nature

| | #Apps | Ads | Analytics | Payment Processing | Crash reporting | User engagement | Deep linking | Authentication | Social Media Integration |
|---|---|---|---|---|---|---|---|---|---|
| com.facebook | 87 | | • | | | | | | • |
| com.squareup | 58 | | • | • | | | | | |
| io.fabric | 44 | | • | | • | • | | | |
| io.branch | 38 | | | | | • | • | | |
| net.hockeyapp | 34 | | | | • | • | | | |
| com.crashlytics | 28 | | • | | • | | | | |
| com.appboy | 26 | | • | | | • | | | |
| com.mopub | 24 | • | | | | | | | |
| com.applovin | 20 | • | • | | | | | | |

## 3.3 Where does the PI leak?

We now discuss each leak destination, quantified in Table 4. *Net* indicates that the information is leaked over the network; these leaks are the most concerning, because the moment the information has left the phone, the user has de facto lost control over where the information propagates. Please note that characterizing the network *domain* where information leaks to over the network – e.g., the app's own servers, or analytics/advertisers – is challenging and requires a dynamic analysis (Wei et al., 2012). *Log* indicates that PI information is leaked to system logs; this is problematic because system logs are visible to any app. *Local DB/Bundles/Shared Pref* indicates that the information is leaked to local persistent storage (*Bundle* is the Android serialization mechanism, and *SharedPreferences* store user preferences). *File/IO* indicates that the PI is saved in local files. Note that leaks to local storage (DB, files) have been proven to be used by malicious apps and malicious libraries to exfiltrate data (Reardon et al., 2019). The table reveals that, depending on the destination, third-party leaks are 2.8x–13.4x more numerous than first-party leaks. For space reasons we omit showing the distribution for each destination, however we found that apps that have first-party network leaks typically have few (<20) whereas 40 apps had ≥ 20 network leaks (they are all third-party leaks).

Table 4. Leak destinations

| | #Leaks | % of overall leaks | First-party | Third-party |
|---|---|---|---|---|
| Net | 7534 | 28.90 | 592 | 6942 |
| Local DB/Bundle/SharedPreferences | 7998 | 30.68 | 1490 | 6508 |
| Log | 8637 | 30.12 | 2272 | 6365 |
| File, I/O | 1903 | 7.30 | 132 | 1771 |

Table 5. Fine-grained PI leak information (#of apps exhibiting *one or more leaks* of that PI)

| | **Fitness** | | | | | **Medical** | | | | |
|---|---|---|---|---|---|---|---|---|---|---|
| | Net | DB | Log | FileI/O | *Total* | Net | DB | Log | FileI/O | *Total* |
| Email | 23 | 67 | 47 | 7 | *144* | 56 | 153 | 124 | 13 | *346* |
| First name | 9 | 6 | 30 | 5 | *50* | 5 | 12 | 7 | 2 | *26* |
| Last name | 12 | 7 | 31 | 5 | *55* | 14 | 11 | 4 | 2 | *31* |
| Phone | 14 | 15 | 9 | 0 | *38* | 20 | 8 | 36 | 3 | *67* |
| Address | 55 | 94 | 28 | 0 | *177* | 163 | 94 | 49 | 1 | *307* |
| Zip | 0 | 0 | 0 | 0 | *0* | 3 | 0 | 0 | 0 | *3* |
| Gender | 7 | 18 | 2 | 0 | *27* | 16 | 17 | 7 | 0 | *40* |
| SSN | 0 | 0 | 0 | 0 | *0* | 0 | 0 | 0 | 0 | *0* |
| CCard | 0 | 0 | 0 | 0 | *0* | 0 | 0 | 2 | 0 | *2* |
| Age | 13 | 10 | 12 | 0 | *35* | 6 | 36 | 6 | 0 | *48* |
| Weight | 93 | 137 | 71 | 28 | *329* | 68 | 47 | 22 | 0 | *137* |
| Height | 39 | 45 | 31 | 17 | *132* | 18 | 13 | 9 | 0 | *40* |
| Medical hist. | 0 | 0 | 0 | 0 | *0* | 62 | 6 | 14 | 2 | *84* |
| Medication | 2 | 35 | 0 | 0 | *37* | 10 | 29 | 11 | 0 | *50* |
| Blood | 0 | 4 | 0 | 0 | *4* | 0 | 7 | 0 | 0 | *7* |
| Mental health | 0 | 0 | 0 | 0 | *0* | 4 | 37 | 4 | 0 | *45* |
| Smoke/Alcohol | 0 | 7 | 0 | 0 | *7* | 0 | 0 | 0 | 0 | *0* |
| *Total* | *267* | *445* | *261* | *62* | *1035* | *447* | *470* | *295* | *30* | *1233* |

## 3.4 Fine-grained PI leak analysis

Table 5 shows fine-grained results: the number of apps in each category that exhibit at least one leak for that PI, as well as totals for each PI and each destination. Note that an app can have multiple leaks of the same PI to the same destination, e.g., the *Email* is saved into two different files, or sent onto the network via two different connections. *In this subsection only, when referring to "leaks" we count apps, not individual leaks; in other words, any app that has more than one PI→destination leak is only counted once.* Overall, Medical apps have about 20% more leaks than Fitness apps (1233 vs. 1035) in part due to medical apps collecting specific medical PI; nevertheless, some Fitness apps still collect and leak medical PI such as medication, blood, or smoking/alcohol use. *Email*, *Address*, *Weight*, and *Height* are the most frequently leaked PI. Interestingly, while 44% of Fitness apps collect the *Email*, only about 14% leak it (see Figure 1); whereas 39% of Medical apps collect the *Email*, and 34% leak it. Another interesting point is that Fitness apps have a much higher tendency to collect and leak *Weight* and *Height*. Finally, Medical apps have 67% more network leaks than Fitness apps do (447 vs. 267), making them higher risk (and inviting more scrutiny) than Fitness apps.

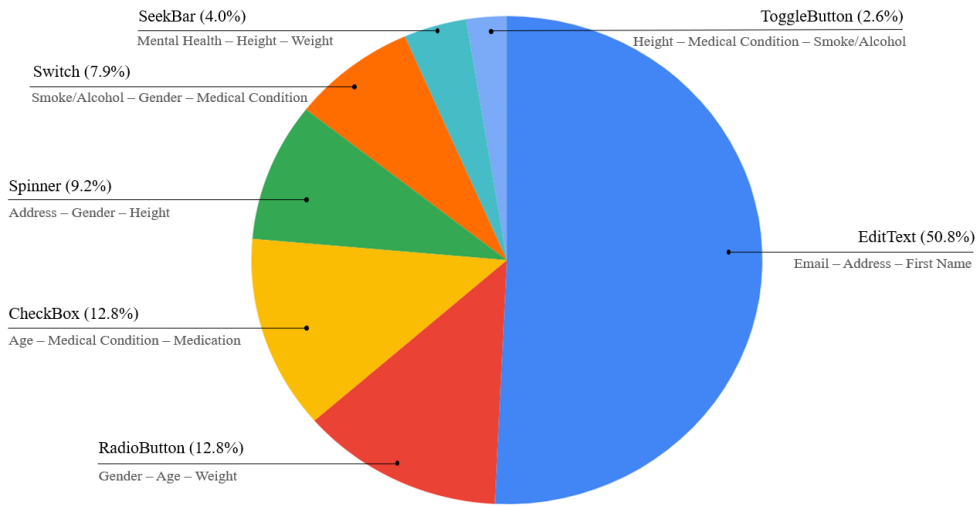

Figure 6. UI element distributions and the PI they collect

## 3.5 Characterizing the UI elements that collect PI

In total, our apps contain 47,749 Android *View* objects (i.e., GUI elements) that collect PI. Figure 6 shows the *View* types, along with the collected PI semantics. For each type, we list the top-3 most frequent collected PI in that *View*. The most prevalent UI object is *EditText* (50.8%) which allows arbitrary text input. Naturally, *EditText* is used to collect emails, addresses, first names, and so on. However, the flexibility of *EditText* can be a downside, when the collected information has a certain type (e.g., numeric) or range, e.g., 0–100; in such cases, developers need to add input validation, whereas other controls, e.g., *Spinner* or *SeekBar*, can directly enforce discipline on values or ranges. *RadioButton* and *CheckBox* are tied for the second most frequent PI collectors, typically used to select the gender, an age range, or a list of medications/medical conditions.

## 3.6 Discussion

We believe that (1) such apps should move toward collecting and leaking less, and (2) users, developers, app markets, and regulators can all play a role. First, users should question developers' over-collection, e.g., an app that simply computes the BMI should just ask for weight and height, and not ask for the user's medical history. Second, as developers typically use third-party code for monetization (ads), they should balance revenue with the leaks that third-party code induces; similarly, users can put pressure on developers to reduce ads and leaks. Third, app markets can be more transparent about the data apps collect, e.g., offer a detailed description of the data collected, and where this data is stored/sent. Fourth, regulators should be much more aggressive in enacting measures to make apps transparent about collection, and protective of user data.

## 4. CONCLUSIONS AND FUTURE WORK

Our study has revealed that medical, health, and fitness apps collect and leak a plethora of personal information. Future work directions include: a dynamic analysis to categorize the destination of PI data; reporting PI data collection and exfiltration for a given individual app to the end-users, e.g., as a browser extension or on-the-phone; informing users of the regulatory environments the app complies with (or conversely, violates), e.g., GDPR in the EU or CCPA in California.

## ACKNOWLEDGMENTS

We thank the anonymous reviewers for their feedback. This material is based upon work supported by the National Science Foundation under Grant No. CCF-2106710.## REFERENCES

Arzt, S. et al. (2014) "FlowDroid: Precise Context, Flow, Field, Object-Sensitive and Lifecycle-Aware Taint Analysis for Android Apps". PLDI '14: Proceedings of the 35th ACM SIGPLAN Conference on Programming Language Design and Implementation.

Continella, A., et al. (2017) "Obfuscation-Resilient Privacy Leak Detection for Mobile Apps Through Differential Analysis" Network and Distributed System Security (NDSS) Symposium 2017.

HIPAA (2024): US Department of Health & Human Services: Guidance Regarding Methods for Deidentification of Protected Health Information in Accordance with the Health Insurance Portability and Accountability Act (HIPAA) Privacy Rule https://www.hhs.gov/hipaa/for-professionals/privacy/special-topics/de identification/index.html#protected.

Huang, J. et al. (2015) SUPOR: precise and scalable sensitive user input detection for android apps. In Proceedings of the 24th USENIX Conference on Security Symposium (SEC'15). USENIX Association, USA.

JADX (2024) Dex to Java decompiler https://github.com/skylot/jadx.

Jia, Q., et al. (2019). "Who Leaks My Privacy: Towards Automatic and Association Detection with GDPR Compliance." In: Biagioni, E., Zheng, Y., Cheng, S. (eds) Wireless Algorithms, Systems, and Applications. WASA 2019. Lecture Notes in Computer Science(), vol 11604. Springer, Cham.

McClurg, J., et al. (2013) "Android Privacy Leak Detection via Dynamic Taint Analysis" Northwestern University https://jrmcclurg.com/papers/internet_security_final_report.pdf.

Rahaman, S., et al. (2021) "Algebraic-datatype taint tracking, with applications to understanding Android identifier leaks." 29th European Software Engineering Conference and Symposium on the Foundations of Software Engineering.

Reardon, J., et al. (2019) "50 Ways to Leak Your Data: An Exploration of Apps' Circumvention of the Android Permissions System" USENIX Security Symposium 2019.

Ren, J., et al. (2016) "ReCon: Revealing and Controlling PII Leaks in Mobile Network Traffic" MobiSys'16: The 14th Annual International Conference on Mobile Systems, Applications, and Services.

Samat, S (2022) Living in a multi-device world with Android. https://blog.google/products/android/io22-multideviceworld/ May 11, 2022.

SuSi (2024). https://github.com/secure-software-engineering/SuSi.

Van Alstin, C (2024) Massive data trove from Change Healthcare hack now for sale on dark web; April 17, 2024 https://healthexec.com/topics/health-it/cybersecurity/massive-data-trove-change-healthcare-hack-now-sale-dark-web.

Wei, X., et al. (2012) ProfileDroid: Multi-layer Profiling of Android Applications. The 18th Annual International Conference on Mobile Computing and Networking (MobiCom 2012), August 2012.